\documentclass[runningheads]{llncs}
\usepackage[T1]{fontenc}
\usepackage{subcaption}
\usepackage{graphicx}
\usepackage{amsmath}
\usepackage{siunitx}
\graphicspath{{images/}}
\usepackage{xcolor}

\usepackage{caption}
\usepackage{glossaries}
\makeglossaries

\newacronym{iot}{IoT}{Internet of Things}
\newacronym{lpl}{LPL}{Low-Power Listening}
\newacronym{wor}{WoR}{Wake-on-Radio}
\newacronym{wur}{WuR}{Wake-up-Radio}
\newacronym{pmic}{PMIC}{Power Management Integrated Circuit}
\newacronym{mppt}{MPPT}{Maximum Power Point Tracking}
\newacronym{mcu}{MCU}{Microcontroller Unit}
\newacronym{gpio}{GPIO}{General-Purpose Input/Output}
\newacronym{fsm}{FSM}{Finite State Machine}
\newacronym{vsp1}{VSP1}{Voltage Supervisor 1}
\newacronym{vsp2}{VSP2}{Voltage Supervisor 2}
\newacronym{ecdf}{ECDF}{Empirical Cumulative Distribution Function}
\newacronym{tdma}{TDMA}{Time-Division Multiple Access}
\newacronym{mac}{MAC}{Medium Access Control}
\newacronym{ti}{TI}{Texas Instruments}
\newacronym{vlc}{VLC}{Visible Light Communication}

\glsenableentrycount
\makeglossaries

\makeatletter
\newcommand{\midsize}{\@setfontsize\midsize{8}{9.0}}
\makeatother

\usepackage{booktabs}
\usepackage{eso-pic}

\newcommand{\AddArxivNotice}{%
  \AddToShipoutPictureFG*{%

    \AtPageUpperLeft{%
      \raisebox{-1.15cm}[0pt][0pt]{%
        \makebox[\paperwidth][c]{%
          \color[gray]{0.58}%
          \fontsize{7}{8.2}\selectfont
          This is an open-access, author-archived version of a manuscript
          published in ApplePies 2026 Conference.
        }%
      }%
    }%
  }%
}

\begin{document}
\AddArxivNotice
\title{Adaptive Wake-on-Radio Communication for Batteryless IoT Systems}

\author{Hooman Sarvghadi\inst{1}
\and
Juan Felipe Gutiérrez Gómez\inst{1,2}
\and
Negar Halakou\inst{1}
\and
Nasrin Razmi\inst{1}
\and
Andrés Gómez\inst{1}
}
\authorrunning{H. Sarvghadi et al.}
\institute{
 Institute of Computer and Network Engineering, Technical University of Braunschweig, Hans-Sommer-Str. 66, Braunschweig, 38106, Germany 
 \and
School of Engineering, Department of Electrical and Electronic
Engineering, Universidad Nacional de Colombia, Ave Cra 30 No. 45-3, Bogotá, 111321, Colombia
\email{{\{hooman.sarvghadi, juan-felipe.gutierrez-gomez, negar.halakou, n.razmi, andres.gomez\}@tu-braunschweig.de}}\\
}
%
\maketitle              
\begin{abstract}
\vspace{-0.2 cm}

Batteryless Internet of Things (IoT) systems powered by energy harvesting require communication mechanisms that dynamically adapt to fluctuating energy availability while maintaining ultra-low-power operation. This paper presents a Wake-on-Radio (WoR) communication framework for low-power intermittent IoT systems in which the node's communication state adapts to its stored energy. The framework uses a batteryless IoT node that integrates a photovoltaic harvester, a \SI{570}{\micro\farad} capacitor, two voltage supervisors, and a microcontroller. The voltage supervisors drive an energy-aware finite state machine that switches the node between Shutdown, Low-Power Listening (LPL), Transmit, and Receive modes. The proposed system employs the integrated WoR capability of the transceiver, eliminating the need for additional wake-up hardware. For the evaluated traffic conditions and an LPL wake-up rate of \SI{16}{\hertz}, the measured power consumption is approximately \SI{1.2}{\micro\watt} in the Shutdown mode, \SI{300}{\micro\watt} during LPL, \SI{531}{\micro\watt} for reception, and \SI{627}{\micro\watt} for transmission. System-level experiments under solar-harvesting operation validate the proposed framework, demonstrating reliable bidirectional communication with no packet losses observed across repeated runs at a harvested input power of \SI{1.2}{\milli\watt}.
\vspace{-0.1 cm}
\vspace{-0.3 cm}
\end{abstract}
\vspace{-0.5 cm}
\section{Introduction}
\vspace{-0.1 cm}
In batteryless systems, sensing and communication can only be performed when sufficient energy has been harvested and stored in a local energy buffer~\cite{gomez2020demand}. In such systems, communication is particularly challenging because device activity is driven by harvested energy rather than by a predetermined schedule. As a result, nodes wake up, transmit, and receive at unpredictable times depending on their available energy. Successful wireless communication requires overlap between transmission and receiver availability, which is difficult to guarantee in independently powered batteryless devices~\cite{babatunde2024greentooth}. 

A conventional approach for improving reception probability is to increase receiver availability through longer listening windows or continuous channel monitoring. However, this comes at a substantial energy cost and can significantly limit the operating lifetime of batteryless devices between energy-harvesting cycles. \cgls{wor} addresses this challenge by enabling asynchronous packet detection while maintaining a very low receiver duty cycle. Although WoR improves the likelihood of successful communication, its effectiveness depends on the trade-off between listening frequency, preamble duration, and communication energy cost. This paper investigates this trade-off and experimentally evaluates an energy-adaptive \cgls{wor} communication framework on a \cgls{ti} CC26X2R1 platform for ultra-low-power batteryless \cgls{iot} systems.

\vspace{-0.3 cm}
\section{Background and Related Works}
\vspace{-0.1 cm}
\cgls{wor} is an energy-efficient communication technique in which a receiver remains in a low-power sleep state for most of the time and periodically wakes for brief channel-sniffing intervals. Upon detecting a valid radio signal or preamble, typically through RSSI-based sensing, the receiver stays active to receive the incoming packet~\cite{maharajan2022wake}. To ensure reliable communication in this asynchronous scheme, the transmitter sends a preamble whose duration is at least the receiver wake-up interval, thereby ensuring overlap with a sniffing window and enabling reliable packet detection~\cite{gallego2013smarter}. 

Prior work has studied WoR and \cgls{wur}-based communication for low-power wireless sensor networks and batteryless \cgls{iot} systems. 
In this context, Mosavat et al.~\cite{mosavat2025cato} proposed a comprehensive wake-up receiver (WuR)-based communication framework for batteryless devices, demonstrating the effectiveness of dedicated \cgls{wur} hardware for energy-efficient communication. Their system employs an FH101RF wake-up receiver alongside a separate radio transceiver. Similarly, an asynchronous \cgls{wor}-based \cgls{mac} protocol on \cgls{ti} CC430 hardware used native radio-core \cgls{wor} and overhearing avoidance to achieve a 0.78\% duty cycle and up to 97\% energy reduction, but it did not adapt communication behavior to varying energy availability~\cite{maharajan2022wake}. 
Geissdoerfer et al. ~\cite{geissdoerfer2022learning} introduced Bonito, a communication protocol for intermittently powered batteryless devices that learns their charging-time distributions and coordinates future wake-up times.

For batteryless \cgls{iot} systems, wake-up receiver-based frameworks with supercapacitor storage and adaptive measurement throttling have demonstrated long-term outdoor operation under variable harvested-energy conditions, achieving 47.83--97.36\% system availability~\cite{mosavat2023experience}. Adaptive duty-cycle optimization has also been studied for neighbor discovery under fixed energy budgets using \cgls{wor} and \cgls{wur} techniques~\cite{loreti2019optimized}. Beyond RF-based communication, energy-aware batteryless architectures have also been explored using alternative communication technologies. A batteryless \cgls{vlc}-based \cgls{iot} node has been proposed where a solar panel operates simultaneously as an energy harvester and optical receiver, enabling communication tasks to be coordinated with the energy-harvesting process~\cite{gutierrez2025solar}. Nevertheless, the wake-up receiver-based approaches described
above~\cite{mosavat2025cato,mosavat2023experience} depend on dedicated hardware alongside the main
transceiver, and Babatunde et al.~\cite{babatunde2024greentooth} similarly combine
low-power wake-up radios with TDMA-style scheduling. Bonito~\cite{geissdoerfer2022learning} avoids
additional hardware but requires a learned timing model and inter-node
coordination that must survive power failures. This work instead uses the
integrated on-chip \cgls{wor} capability of a single transceiver for both \cgls{lpl} and
packet exchange, and selects communication states locally from the
stored-energy level, without dedicated hardware, inter-node
synchronization, or continuous voltage sampling.

\vspace{-0.4 cm}
\section{System Design and Implementation}
\begin{figure}[t]
    \centering

    \begin{subfigure}[t]{0.43\textwidth}
        \centering
        \includegraphics[width=\linewidth]{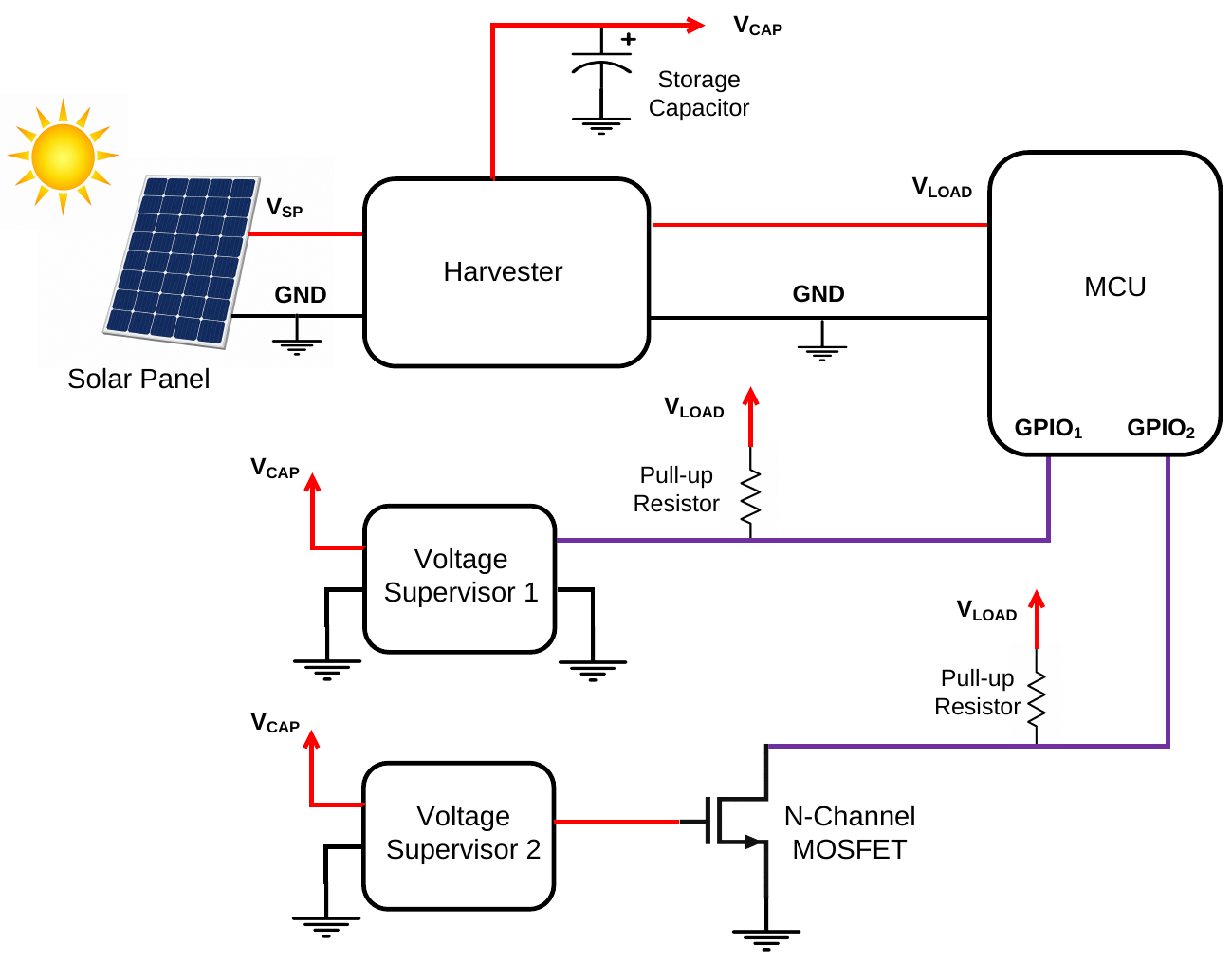}
        \caption{Hardware architecture}
        \label{fig:system_architecture}
    \end{subfigure}
    \hspace{0.055\textwidth}
    \begin{subfigure}[t]{0.31\textwidth}
        \centering
        \includegraphics[width=\linewidth]{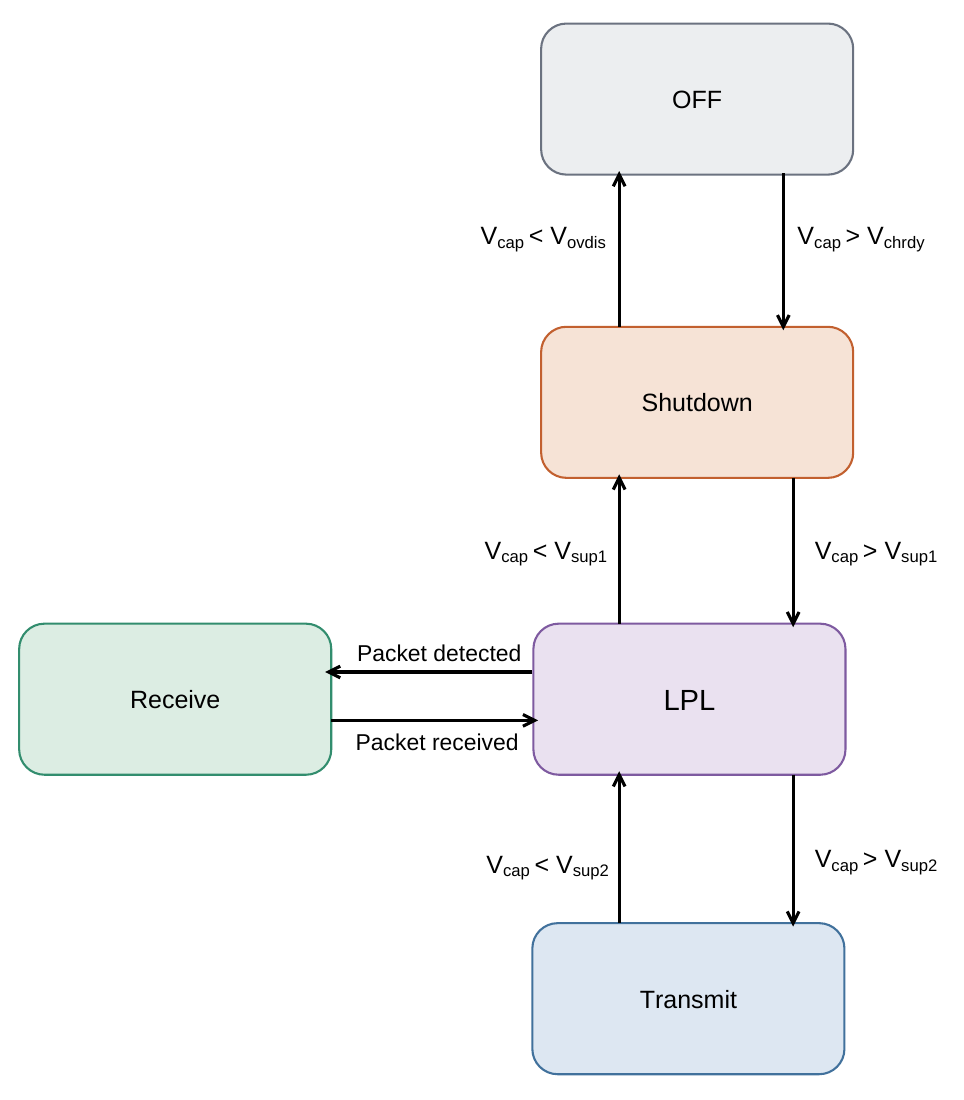}
        \caption{Finite State Machine (FSM)}
        \label{fig:fsm}
    \end{subfigure}

    \caption{\scriptsize
    Overview of the proposed system.
    (a) Hardware architecture of the energy-harvesting node.
    (b) FSM describing the energy-adaptive operating modes.
    \vspace{-0.6 cm}}
    \label{fig:hardware_architecture}
\end{figure}

\vspace{-0.05 cm}
We consider a batteryless \cgls{iot} node powered by harvested solar energy and buffered by a capacitor. Since the stored energy varies with harvesting conditions, the node must detect its energy state and adapt communication with minimal processing overhead. The proposed hardware framework is illustrated in Fig.~\ref{fig:hardware_architecture}(a). It consists of a photovoltaic panel, an e-peas AEM10941 energy-harvesting evaluation board, TPS3840DL35DBVR and TPS3840PL43DBVR voltage supervisors with \SI{200}{\milli\volt} hysteresis, a \SI{570}{\micro\farad} capacitor, and a \cgls{ti} CC2652R1 LaunchPad as the \cgls{mcu}. The AEM10941 performs \gls{mppt}, charges the capacitor, and supplies a regulated voltage to the \cgls{mcu}. The capacitor voltage $V_{\mathrm{cap}}$ is monitored by both supervisors, whose outputs are connected to two \glspl{gpio} through \SI{1.2}{\mega\ohm} pull-up resistors, with an N-channel MOSFET in the signal path of the second supervisor.

The proposed \cgls{fsm}, shown in Fig.~\ref{fig:hardware_architecture}(b), adapts the communication mode according to stored energy, using voltage-supervisor thresholds, while the \cgls{lpl} wake-up rate is fixed at design time. The node remains OFF while the capacitor charges and, when $V_{\mathrm{cap}}>V_{\mathrm{chrdy}}$, enters Shutdown mode. When $V_{\mathrm{cap}}>V_{\mathrm{sup1}}$, it enters \cgls{lpl}, periodically performing \cgls{wor} channel sniffing; if $V_{\mathrm{cap}}<V_{\mathrm{sup1}}$, it returns to Shutdown for recharging. In \cgls{lpl}, $V_{\mathrm{cap}}>V_{\mathrm{sup2}}$ triggers a packet transmission, after which the node returns to \cgls{lpl}. Detection of an incoming packet similarly switches the node to receive mode before returning to \cgls{lpl}. If $V_{\mathrm{cap}}<V_{\mathrm{ovdis}}$, the system enters OFF mode until $V_{\mathrm{cap}}>V_{\mathrm{chrdy}}$. This hardware-assisted control provides energy-aware communication without continuous \cgls{mcu} voltage sampling, reducing computational and energy overhead.

\vspace{-0.4 cm}
\section{Results and Discussion}
\vspace{-0.1 cm}
We evaluate the proposed framework by first quantifying the communication energy cost for different wake-up rates. This characterization guides the wake-up configuration that minimizes node power under each harvesting condition. The wake-up rate is varied over 2, 4, 8, 16, and 32~Hz, while all other communication parameters fixed. Experiments are conducted with a 255-byte payload at an input voltage of 2.5~V. Each measurement lasts $T = 300$~s, during which the transmitter generates $N_{\mathrm{pkt}} = 52$ communication events. Across repeated runs, all transmitted packets were correctly received by the peer, with no losses observed. For each wake-up configuration, we measure the average \cgls{lpl} power and the energy per transmitted and received packet. Table~\ref{tab:wps_comparison} summarizes the measured packet-level energy and average background power for the wake-up rate configurations. Higher wake-up rates reduce packet-level energy by shortening the required preamble duration, while increasing the average \cgls{lpl} power due to more frequent channel-sniffing.

\begin{table}[t]
\caption{\scriptsize{Measured node power and energy across wake-up rates.}}
\label{tab:wps_comparison}
\centering
\footnotesize
\setlength{\tabcolsep}{7pt}
\resizebox{0.75\columnwidth}{!}{%
\begin{tabular}{lcccccc}
\toprule
&
\multicolumn{2}{c}{\textbf{Per-Event Energy}} &
\multicolumn{2}{c}{\textbf{LPL Power}} &
\multicolumn{2}{c}{\textbf{Whole-Capture Node Power}}\\
\cmidrule(lr){2-3}
\cmidrule(lr){4-5}
\cmidrule(lr){6-7}

$\mathbf{WR}$ [Hz] &
$\mathbf{E_{\mathrm{TX}}}$ [$\mu$J] &
$\mathbf{E_{\mathrm{RX}}}$ [$\mu$J] &
$\mathbf{P_{\mathrm{LPL}}^{\mathrm{TX}}}$ [$\mu$W] &
$\mathbf{P_{\mathrm{LPL}}^{\mathrm{RX}}}$ [$\mu$W] &
$\mathbf{P_{\mathrm{Avg}}^{\mathrm{TX}}}$ [$\mu$W] &
$\mathbf{P_{\mathrm{Avg}}^{\mathrm{RX}}}$ [$\mu$W] \\
\midrule

2  & 10914 & 5222 & 32.71 & 39.99 & 1875 & 1015 \\
4  & 5497  & 3052 & 69.13 & 74.17 & 1094 & 645 \\
8  & 2965  & 1670 & 142.03 & 148.72 & 714 & 472 \\
16 & 1700  & 1131 & 294.89 & 309.27 & 627 & 531 \\
32 & 1062 & 786 & 662.23 & 712.63 & 872 & 868 \\

\bottomrule
\end{tabular}
}\vspace{-0.5 cm}
\end{table}

\begin{figure}[t]
    \centering

    \begin{subfigure}[t]{0.365\textwidth}
        \centering
        \includegraphics[width=\linewidth]{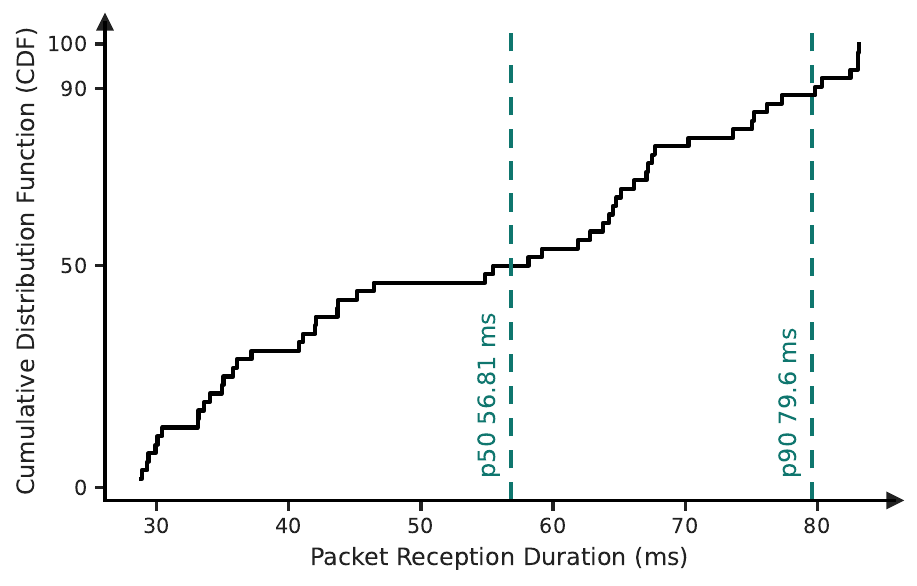}
        \caption{}
        \label{fig:rx_duration_cdf}
    \end{subfigure}%
    \hspace{0.002\textwidth}%
    \begin{subfigure}[t]{0.315\textwidth}
        \centering
        \includegraphics[width=\linewidth]{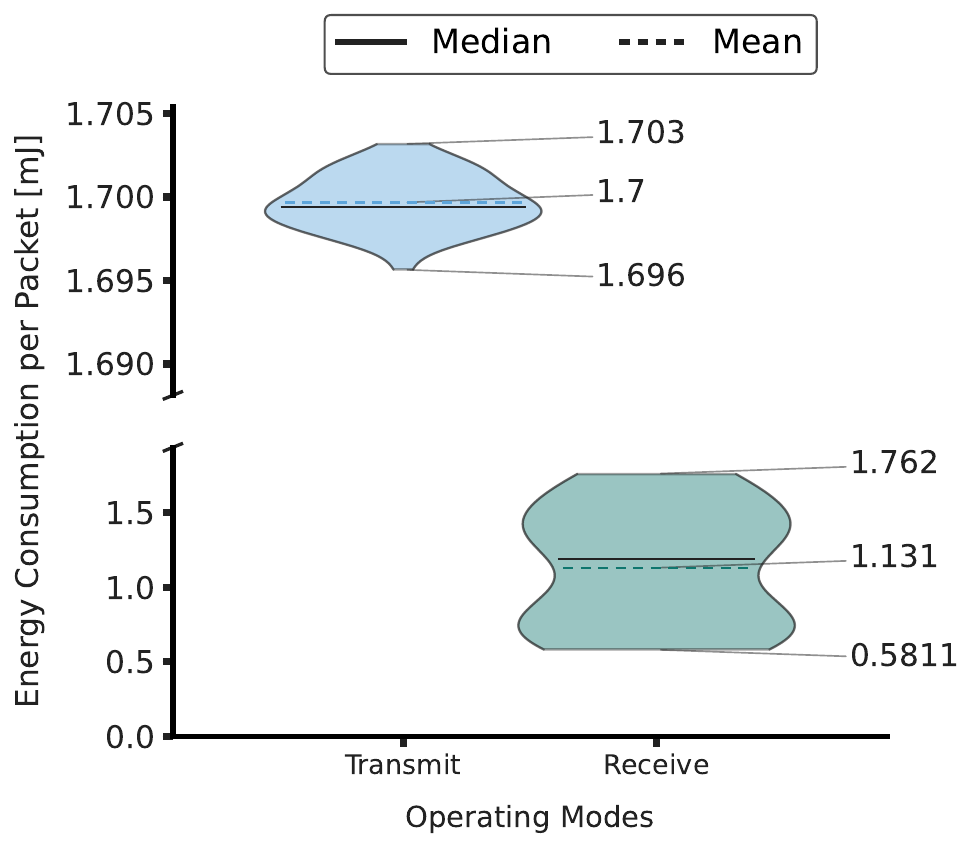}
        \caption{}
        \label{fig:energy_breakdown}
    \end{subfigure}%
    \hspace{0.002\textwidth}%
    \begin{subfigure}[t]{0.315\textwidth}
        \centering
        \includegraphics[width=\linewidth]{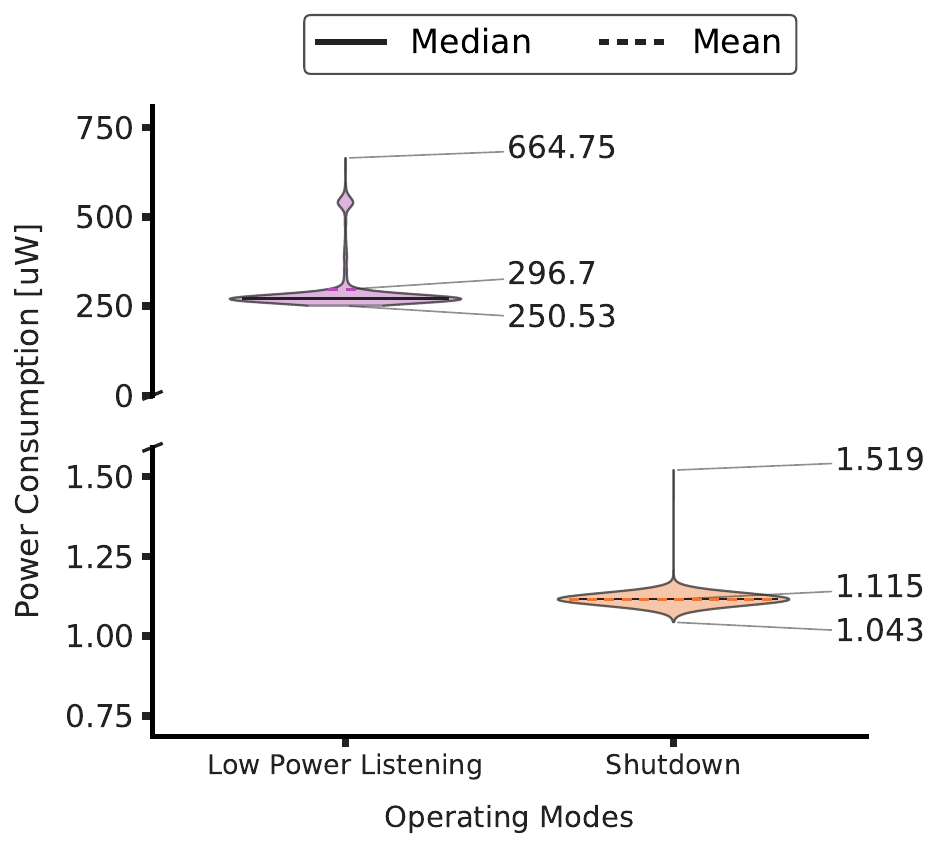}
        \caption{}
        \label{fig:figure_c}
    \end{subfigure}

    \vspace{-0.2cm}

    \caption{\scriptsize{
        (a) Empirical cumulative distribution function (ECDF) of RX duration.
        (b) Measured energy for a transmitted and received packet.
        (c) The steady-state power for \cgls{lpl} and Shutdown modes.
    }\vspace{-0.6cm}}
    
    \label{fig:txrx_lpl_comparison}
\end{figure}

Figures \ref{fig:txrx_lpl_comparison}b and \ref{fig:txrx_lpl_comparison}c present the distributions of the energy characteristics, including packet energy for transmit and receive states and steady-state power for \cgls{lpl} and Shutdown modes. The transmission energy per packet remains nearly constant across measurements, whereas the reception energy shows a wider spread due to variations in packet reception duration. In contrast, \cgls{lpl} mode consumes substantially higher steady-state power than Shutdown mode. Consequently, the optimal wake-up configuration depends not only on the packet-level energy but also on the communication activity, i.e., how frequently packets are transmitted or received and how long the node remains in each steady-state operating mode. The \cgls{ecdf} (ECDF) in Fig.~\ref{fig:rx_duration_cdf} further demonstrates the spread of the reception energy.

\begin{figure}[t]
    \centering

    \begin{subfigure}[t]{0.34\textwidth}
        \centering
        \includegraphics[
            width=\linewidth,
        ]{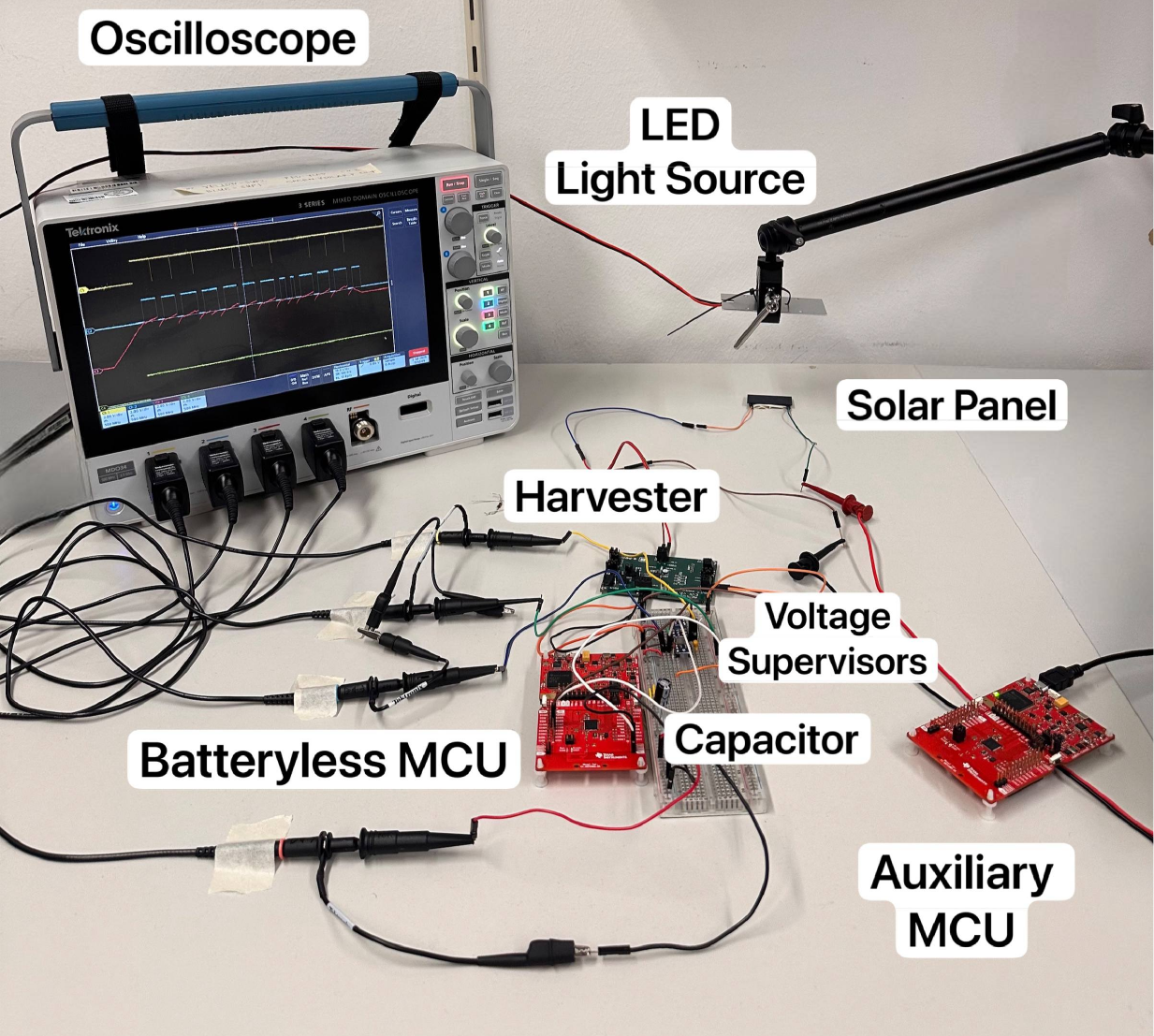}
        \caption{\scriptsize Experimental setup}
        \label{fig:experimental_setup}
    \end{subfigure}
    \hfill
    \begin{subfigure}[t]{0.65\textwidth}
        \centering
        \includegraphics[width=\linewidth]{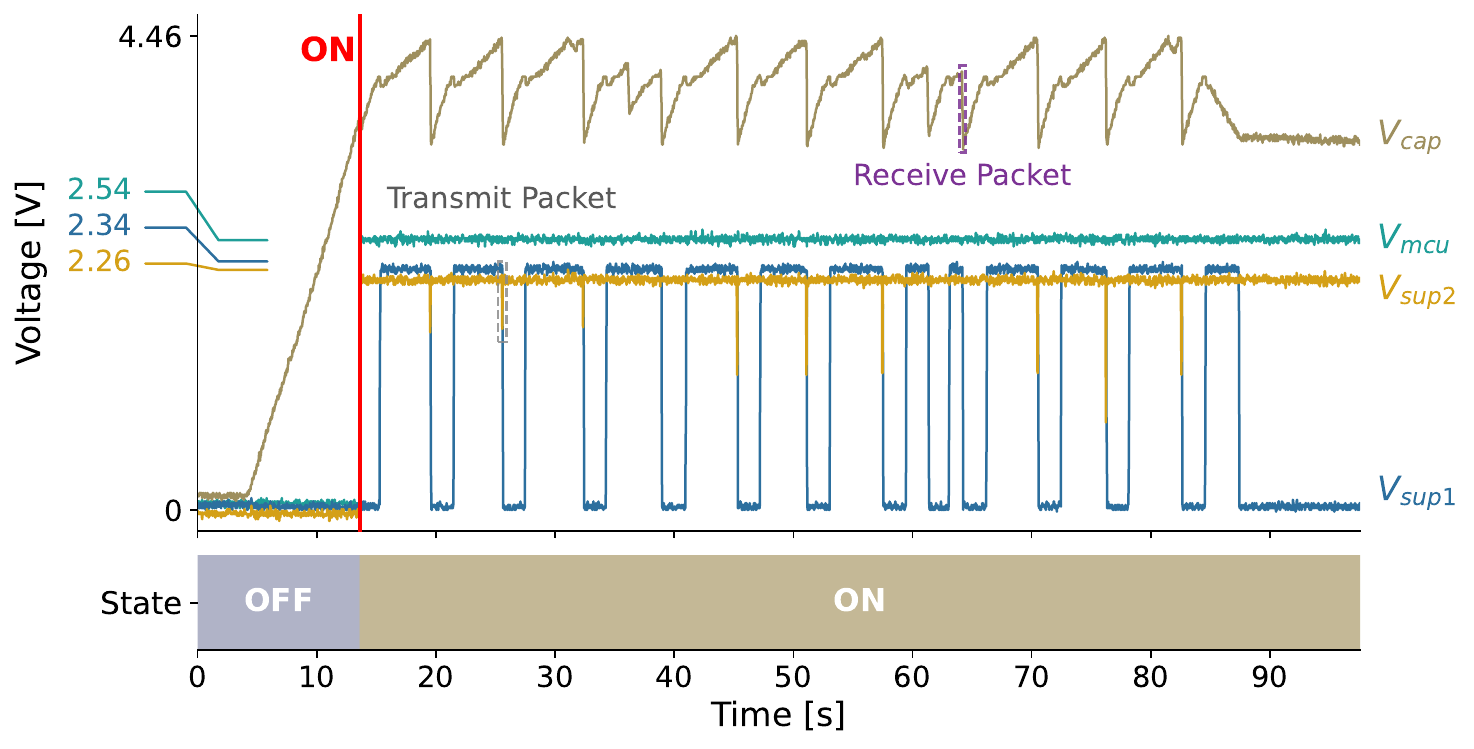}
        \caption{\scriptsize Measured operating profile}
        \label{fig:power_profile}
    \end{subfigure}
\vspace{- 0.2 cm}
    \caption{\scriptsize{Experimental setup and measured voltage signals from the capacitor, voltage supervisor outputs, and load voltage at a solar input power of \SI{1.2}{\milli\watt}.}
    \vspace{-0.4 cm}}
    \label{fig:experiment}
\end{figure}

The system-level experimental setup is shown in Fig.~\ref{fig:experimental_setup}. The complete batteryless communication platform is powered solely by a photovoltaic panel, while an oscilloscope monitors the capacitor voltage, regulated supply voltage, voltage supervisor outputs, and communication events during system operation. An auxiliary \cgls{mcu} running the same firmware acts as the peer, transmitting packets and verifying successful packet reception by the batteryless \cgls{mcu}. Fig.~\ref{fig:power_profile} shows the end-to-end operation of the proposed batteryless energy-adaptive communication system powered by a solar panel, under illumination conditions that provide approximately \SI{1.2}{\milli\watt} of harvested power. Initially, the storage capacitor charges until its voltage \(V_{\mathrm{cap}}\) reaches the harvester start-up threshold. At this point, the regulated \cgls{mcu} supply voltage \(V_{\mathrm{mcu}}\) is enabled, and the system powers on. After power-up, the node enters Shutdown mode to conserve energy while the capacitor continues charging. As \(V_{\mathrm{cap}}\) exceeds the threshold of $V_{\mathrm{sup1}}$, the voltage supervisor output is driven high and the node enters \cgls{lpl} mode, where the radio periodically samples the channel. When \(V_{\mathrm{cap}}\) further exceeds the threshold of $V_{\mathrm{sup2}}$, the second voltage supervisor activates, \(V_{\mathrm{sup2}}\) is asserted, indicating sufficient energy for transmission. The node then transmits a packet and returns to \cgls{lpl} mode. Between transmissions, multiple packet receptions are observed, demonstrating bidirectional communication powered solely by harvested energy, with all transmitted packets correctly received across the repeated runs. Each communication event produces a corresponding discharge of the storage capacitor, visible as a temporary drop in \(V_{\mathrm{cap}}\). Following every transmission or reception, harvested solar energy recharges the capacitor until the next communication event, resulting in characteristic charge--discharge cycles. Finally, when the solar panel is disconnected, harvesting stops and the capacitor gradually discharges until the available energy is insufficient to sustain system operation.

Throughout the experiment, the regulated \cgls{mcu} supply voltage 
remains nearly constant despite large variations in \(V_{\mathrm{cap}}\), demonstrating stable \cgls{mcu} operation while communication adapts to the available harvested energy. One limitation of the proposed framework is that higher harvested input power makes the storage capacitor recharge more rapidly after each transmission. Consequently, the node may spend less time in the \cgls{lpl} mode before the next transmission event. This reflects a threshold-design tradeoff between transmission activity and receiver availability. Future work should jointly optimize voltage thresholds and capacitor sizing to improve receiver availability under varying harvested power conditions.

\vspace{-0.4 cm}
\section{Conclusions}
\vspace{-0.2 cm}
This paper presented an energy-adaptive \cgls{wor} framework for batteryless \cgls{iot} nodes based on the TI CC2652R1 platform. The proposed framework employs a harvested-energy-driven \cgls{fsm} algorithm to switch between Shutdown, \cgls{lpl}, Transmit, and Receive modes without requiring continuous voltage sampling. System-level experiments powered solely by a photovoltaic panel demonstrated stable operation and bidirectional communication under fluctuating harvested energy. Future work will investigate adaptive wake-up rate control based on harvested energy to better balance transmission and receiver availability, and extend the framework to multi-node batteryless \cgls{iot} networks.

\vspace{-0.4 cm}
\section*{Acknowledgment}
\vspace{-0.2 cm}
The authors thank Bhakti Patil for contributions to an earlier version. Claude was used only for minor language edits and limited figure graphics.

\vspace{-0.4 cm}

{
}
\end{document}